\documentclass[twocolumn]{svjour3}          % twocolumn

\usepackage{amsmath,amssymb}
\usepackage{array}
\usepackage{color}
\usepackage{flushend}
\usepackage{graphicx}
\usepackage{hyperref}
\usepackage{multirow}

\usepackage{draftwatermark}
\SetWatermarkText{preprint}
\SetWatermarkScale{1}

\newcommand{\totalN}{41}
\newcommand{\ab}[1] {\textcolor{black}{#1}}
\newcommand{\js}[1] {\textcolor{black}{#1}}
\newcommand{\oz}[1] {\textcolor{black}{#1}}
\newcommand{\abb}[1] {\textcolor{black}{#1}}
\newcommand{\ozz}[1] {\textcolor{black}{#1}}

\begin{document}

% OZ (0315): added "s" to "refactorings"
\title{Design principles, architectural smells and refactorings for microservices: A multivocal review}

\titlerunning{Design principles, architectural smells and refactorings for microservices}

\author{%
    Antonio Brogi \and
    Davide Neri \and
    Jacopo Soldani \and 
    Olaf Zimmermann
}

\authorrunning{Brogi A, Neri D, Soldani J, Zimmermann O}

\institute{%
    Antonio Brogi, Davide Neri, Jacopo Soldani
    \at
    University of Pisa, Pisa, Italy.
    \texttt{\{name\}.\{surname\}@unipi.it}
    \and
    Olaf Zimmermann
    \at
    University of Applied Sciences of Eastern Switzerland (HSR FHO), Rapperswil, Switzerland.
    \texttt{ozimmerm@hsr.ch}
}

\date{Received: date / Accepted: date}

\maketitle

\begin{abstract}
%are pervading enterprise IT, as their potential benefits 
\ozz{Potential benefits such as agile service delivery} \abb{have} led many companies to deliver their  business capabilities through microservices. Bad smells are however always around the corner, as witnessed by the considerable body of literature discussing architectural smells 
\abb{that possibly violate} the design principles of microservices.
In this paper, we systematically review the white and grey literature on the topic, in order to identify \abb{the most recognised} architectural smells for microservices and to discuss the architectural refactorings allowing to resolve them.

%\oz{can we pls change kw from "Architectural style" to "Service-oriented computing" or "SOA"?} 
\end{abstract}

\section{Introduction}
\label{sec:intro}
% OZ: reworded a bit for consistency with my "Tenets" article, feel free to undo. we can also discuss on a call if needed. 
Microservices  architectures, first discussed by Lewis and Fowler~\cite{o2-microservices-fowler}, bring various advantages such as ease of deployment, resilience, and scaling \cite{b1-building-microservices}. 
Many IT companies deliver their core business through microservice-based solutions nowadays\abb{, with} Amazon, Facebook, Go\-ogle, LinkedIn, Netflix and Spotify being prominent examples.
To deliver on their promises, microservices must be designed in quality and style, which is unfortunately not always the case~\cite{m26-microservices-pains-gains}. % oz (0315): replace "well" with "in quality and style" (not urgent/critical so just added as comment for later consideration)

% oz (0315): minor rewordings
Microservice-based architectures can be seen as peculiar extensions of service-oriented architectures, characterized by an extended set of design principles~\cite{Pautasso:2017:MSIP,microservices-tenets-m6}.
These principles include shaping services around business concepts, decentralising all development aspects of microservice-based solutions (from governance to data management), adopting a culture of automation, ensuring the independent deployability  and high observability of microservices, and isolating failures~\cite{b1-building-microservices}.
Key questions are \abb{hence}: 

\begin{quote}
\textit{How to check whether one or more microservices design principles are violated in an application?} %How to resolve potential issues?
\end{quote}
\begin{quote}
\abb{\textit{Which refactorings can be applied to resolve violations of the principles?}}
\end{quote}

The currently available information on architectural smells indicating possible violations of the design principles of microservices is scattered over a considerable amount of literature.
Our objective here is \ab{to} systematically analyse such literature, in order to identify 
\abb{the most recognised} smells, as well as architectural refactorings for \js{resolv}ing the smells occurring in an application \cite{pn-arch-ref-cloud}.
In particular, we focus on the design principles dealing with the dynamic aspects of the interactions between microservices at runtime, i.e.,~on the process viewpoint, as per the 4+1 viewpoint scheme~\cite{4plus1viewpoints}.
More precisely, we consider the independent deployability of microservices, their horizontal scalability, isolation of failures and decentralisation.

As recommended by Garousi et al.~\cite{need-multivocal-reviews}, to capture both the state of the art and the state of practice in the field, we conducted a multivocal systematic review of the existing literature, including both white literature (i.e.,~peer-reviewed papers) and grey literature (i.e.,~blog posts, industrial whitepapers and books).
We selected $\totalN$ studies, published since 2014 (when the microservice-based architectural style was first discussed~\cite{o2-microservices-fowler}) until the end of January 2019.
Then, following the guidelines for systematic reviews~\cite{need-multivocal-reviews,petersen-systematic-reviews}, we excerpted a taxonomy of design principles, architectural smells and corresponding refactorings.
We then exploited \oz{this} taxonomy to classify the selected studies, in order to distill the actual recognition of the identified smells and the usage of the corresponding refactorings.

In this paper, we illustrate the results of our study. 
More precisely, we first present the obtained taxonomy, including \js{seven} architectural smells and 16 refactorings, organised by design principles.
We then discuss each smell, by illustrating why it \abb{can violate} the design principle it \ab{is associated with}, and by showing how to \js{resolve it} by means of an architectural refactoring.

We believe that the results presented in this study can provide benefits to both researchers and practitioners interested in microservices.
A systematic presentation of the state of the art and practice on architectural smells and refactorings for microservices provides a body of knowledge to develop new theories and solutions, to analyse and experiment research implications, and to establish future research directions. 
At the same time, it can help practitioners to \ab{better} understand the currently
\abb{most recognised} architectural smells for microservices, and to choose among the architectural refactorings allowing to \js{resolve} such smells.
This can have a pragmatic value for practitioners, who can use our study as a starting point
for microservices experimentation or as a guideline for day-by-day work with microservices.

% OZ: entire intro reads very well and makes scope and approach of paper clear. some discussion points (for a call?): 
% - anti pattern vs. smell (not all microservices authors use the (anti) pattern metaphor)
% - selection of principles/tenets: maybe give some more rationale (?)
The rest of the paper is organised as follows. 
Sect.~\ref{sec:selection} defines the research problem and illustrates the research methodology.
Sect.~\ref{sec:taxonomy} presents a taxonomy for design principles, architectural smells and refactorings, which is retaken in Sect.~\ref{sec:antipatterns-refactorings} to overview the current state of the art and practice on such smells and refactorings. 
Sects.~\ref{sec:ttv} and \ref{sec:related} discuss potential threats to the validity of our study 
\abb{and related work}, respectively.
Finally, Sect.~ \ref{sec:conclusions} draws some concluding remarks.

\section{Setting the stage}
\label{sec:selection}
\subsection{Research problem definition}
This survey focuses on the architectural principles of microservices that pertain to the process viewpoint, i.e.,~dealing with the dynamic aspects of microservices interacting at 
\ab{runtime~\cite{4plus1viewpoints}}.
%AB I have removed the parenthsis "(e.g.,~integration, failures and scalability)" as it was only mentioning some of aspects of the principles enumerated a few lines below
% oz (0315): min edits
Starting from the principles proposed by Newman~\cite{b1-building-microservices} and by Lewis and Fowler~\cite{o2-microservices-fowler}, and considering the mapping to tenets proposed by Zimmermann ~\cite{microservices-tenets-m6}, \js{we identified} four such principles: 
\begin{itemize}
    \item The microservices forming an application should be \textit{independently deployable}.
    \item The microservices forming an application should be \textit{horizontally scalable}.
    \item \textit{Failures} should be \textit{isolated}.
    \item \textit{Decentralisation} should occur in all aspects of micro\-service-based applications, from data management to governance.
\end{itemize}
%Given the above principles, t
\ab{T}he objective of this survey is to identify the architectural smells indicating  possible violations of such principles, as well as the currently available solutions for refactoring microservice-based architectures in order to \js{resolve} the identified smells.

%OZ (round 2):
Due to space limitations, we cannot cover all microservices tenets from the literature in this paper. Hence, we decided to focus on four particularly relevant principles, driven by three selection criteria: 
\begin{enumerate}
    \item \textit{Roots} in highly significant design time and runtime quality attributes and style-defining elements,  
    \item \textit{Consequences} of not adhering to a principle in terms of technical risk and re-engineering cost, and 
    \item \textit{Generality}, i.e., if these four principles are met, others follow or can be achieved with similar means.
\end{enumerate}
For instance, independent deployability is a defining tenet in most (if not all) definitions of microservices and enables decentralized continuous delivery, thereby meeting criteria (1) and (3). Scalability is a quality attribute (1) and \ozz{horizontal scalabilty is} hard to retrofit (an aspect of (2)). Failure isolation meets criteria (1) and (2). Finally, decentralization is mentioned as crucial (and novel) in many introductions to microservices and enables independent, autonomous decision making, as required to achieve (1) and (3).

%AB: I have reomove the last sentece "As part of our future work, we consider to inspect additional principles." as it can go directly in the conclusions section

% we could also try to draw a simple figure that shows that independent. deployability is a defining prerequisite, which enables the other three  that address the R, the P and the S in FURPS (https://en.wikipedia.org/wiki/FURPS)

\subsection{Search for studies}
\js{With the objective of capturing the state of the art and practice in the field,}
%As recommended by Garousi et al.~\cite{need-multivocal-reviews}, to capture both the state of the art and the state of practice in the field, 
we searched for both white literature (i.e.,~peer-reviewed journal and conference articles) and grey literature (i.e.,~blog posts, industrial whitepapers and books), \abb{in line} with what recommended by Garousi et al.~\cite{need-multivocal-reviews}.
% OZ: repeat from Section 1? 

The structuring of the search string was done by following the guidelines provided by Petersen et al.~\cite{petersen-systematic-reviews}.
We indeed identified the search string guided by the PICO terms of our resarch problem, and the keywords were taken from each aspect of our research problem.
Differently from Petersen et al.~\cite{petersen-systematic-reviews}, we did not restrict our focus to specific research settings.
By restricting ourselves to certain types of research settings, we could have obtained a biased or incomplete analysis, as some architectural smells or refactorings might have been over-/under-represented for a certain type of study.

As a result, our search string was formed by the following terms:
\begin{center}
\ttfamily 
    microservice*
    \\
    $\wedge$
    \\
    (smell* $\vee$ antipattern* $\vee$ bad practice* $\vee$ pitfall* $\vee$ refactor* $\vee$ reengineer*)
\end{center}
(where `\texttt{*}' matches lexically related terms).
The search was restricted to studies published since the beginning of 2014 (when microservices were first proposed by Lewis and Fowler~\cite{o2-microservices-fowler}) until the end of January 2019 (when the present study was \ozz{initiated}). 

% OZ: have you tried "anti pattern" and "anti-pattern" ;-) [just anticipating some potential review comments]? 

The search of white literature was carried out in the following indexing databases: ACM Digital Library, DBLP, EI Compendex, IEEE Xplore, INSPEC, ISI Web of Science, Science Direct, SpringerLink. 
Given the recency of the field and concerns with indexing, Google Scholar played a key role for the initial selection before the inclusion and exclusion stage.
The search for industrial studies was instead carried out in renowned blogs in the software engineering community (such as DZone, InfoQ and TechBeacon), \ab{in} the blog of ThoughtWorks, and \ab{in} books published by practitioners. %  (where microservices were first proposed~\cite{o2-microservices-fowler})

% OZ: FYI: I listed some practitioner sources in a side bar in http://ieeexplore.ieee.org/stamp/stamp.jsp?arnumber=7725214 

\subsection{Sample selection}
The above described search criteria were matched by more than 150 studies, which we carefully screened to keep only those studies that were satisfying  both the following inclusion criteria:
%\footnote{\abb{***REMOVE this footnote?***} Both criteria were checked by verifying whether a study \ab{discusses} the problems characterised by an architectural smell, and whether it \ab{discusses} the architectural changes characterising a refactoring.}:

\begin{itemize}
    \item A study is to be selected if it presents {\em at least one architectural smell} pertaining to one of the considered architectural principles of microservices (i.e., independent deployability, horizontal scalability, isolation of failure, or decentralisation).
    \item A study is to be selected it it presents {\em at least one refactoring} for resolving one of the architectural smells it discusses.
\end{itemize}

The inclusion criteria were defined with the ultimate goal of selecting only representative studies, discussing both the architectural smells (pertaining to the process viewpoint) and their corresponding refactorings.

As a result, $\totalN$ %of the matching
studies were selected to be analysed further. 
The list of references to the selected studies can be found in Table~\ref{tab:selected-studies}, which also classifies them by colour, contribution type and year of publication.

\begin{table}
\centering
\begin{tabular}{c@{\hspace{.2in}}c@{\hspace{.2in}}c@{\hspace{.2in}}c}
    \hline
    \textbf{Ref.} &
        \textbf{Colour} &
        \textbf{Type} &
        \textbf{Year} \\
    \hline
    \cite{o4-seven-microservices-antipatterns} & 
        grey &
        blog post &
        2015 \\
    \cite{m19-systematic-mapping-study-microservice-architecture} &
        white &
        conference & 
        2016 \\ 
    \cite{m23-microservices-devops} & 
        white &
        journal &
        2016 \\ 
    \cite{m25-microservices-migration-patterns} & 
        white &
        journal &
        2018 \\ 
    \cite{o10-design-patterns-microservices} &
        grey &
        blog post & 
        2018 \\ 
    \cite{b9-reactive-microservices} & 
        grey & 
        book &
        2016 \\
    \cite{b5-microservices-from-day-one} & 
        grey &
        blog post & 
        2016 \\ 
    \cite{b11-spring-microservices} &
        grey &
        book &
        2017 \\
    \cite{migrating-towards-microservices-m13} & 
        white &
        conference &
        2018 \\ 
    \cite{o7-performance-patterns-microservices} & 
        grey &
        blog post &
        2016 \\ 
    \cite{m24-architecting-microservices} & 
        white &
        journal &
        2019 \\ 
    \cite{research-on-microservices-m17} & 
        white &
        conference &
        2017 \\ 
    \cite{microservices-yesterday-today-tomorrow-m9} & 
        white & 
        conference &
        2017 \\
    \cite{migrating-enterprise-source-code-to-microservices-m8} &
        white &
        journal &
        2018 \\ 
%    \hline
%\end{tabular}
%\hfill
%\begin{tabular}{c@{\hspace{.2in}}c@{\hspace{.2in}}c@{\hspace{.2in}}c}
%    \hline
%    \textbf{Ref.} &
%        \textbf{Colour} &
%        \textbf{Type} &
%        \textbf{Year} \\
%    \hline
    \cite{o6-develop-great-microservices} & 
        grey &
        blog post &
        2018 \\ 
    \cite{o13-fundamentals-microservice-design} &
        grey &
        blog post &
        2017 \\
    \cite{o11-creating-microservice}  & 
        grey &
        blog post &
        2018 \\ 
    \cite{o8-microservices-in-practice}  & 
        grey &
        blog post &
        2016 \\ 
    \cite{b6-microservices-for-enterprise}  & 
        grey &
        book &
        2018 \\ 
    \cite{microservices-journey-challenges-m7} &
        white &
        journal &
        2018 \\ 
    \cite{m27-challenges-monolith-microservices} & 
        white &
        conference &
        2018 \\ 
    \cite{m11-microservices-for-legacy-software} & 
        white &
        journal &
        2018 \\ 
    \cite{b10-microservices-patterns-applications} & 
        grey & 
        book &
        2015 \\
    \cite{o2-microservices-fowler} & 
        grey & 
        blog post & 
        2014 \\ 
    \cite{o9-power-patterns-microservices} & 
        grey &
        blog post &
        2015 \\ 
    \cite{o12-container-patterns} & 
        grey & 
        blog post & 
        2018 \\ 
    \cite{b4-microservice-architecture} & 
        grey & 
        book &
        2016\\ 
    \cite{b1-building-microservices} & 
        grey & 
        book &
        2015 \\
%    \hline
%\end{tabular}
%\hfill
%\begin{tabular}{c@{\hspace{.2in}}c@{\hspace{.2in}}c@{\hspace{.2in}}c}
%    \hline
%    \textbf{Ref.} &
%        \textbf{Colour} &
%        \textbf{Type} &
%        \textbf{Year} \\
%    \hline
    \cite{b7-release-it} &
        grey &
        book &
        2018 \\
    \cite{b3-microservices-anti-patterns-pitfalls} & 
        grey & 
        book &
        2016 \\ 
    \cite{o1-microservices-intro} & 
        grey & 
        blog post &
        2014 \\ 
    \cite{o3-microservices-io} & 
        grey & 
        book &
        2018 \\ 
    \cite{o14-common-pitfalls} &
        grey &
        blog post &
        2017 \\
    \cite{o5-microservices-antipatterns} & 
        grey & 
        blog post &
        2016 \\ 
    \cite{microservices-mjolnirr-case-study-m14} & 
        white &
        conference &
        2015 \\ 
    \cite{m26-microservices-pains-gains} & 
        white & 
        journal &
        2018 \\ 
    \cite{microservice-bad-smells-m5} & 
        white & 
        journal &
        2018 \\ 
    \cite{processes-motivations-issues-microservices-m16} &
        white & 
        journal &
        2017 \\ 
    \cite{architectural-patterns-microservices-m3} & 
        white & 
        conference &
        2018 \\ 
    \cite{b8-microservices-flexible-sw-architecture} & 
        grey & 
        book &
        2016 \\ 
    \cite{microservices-tenets-m6} & 
        white & 
        journal &
        2017 \\
    & & & \\
    \hline
\end{tabular}
\caption{References to the selected studies, and their classification by \textit{colour} (i.e.,~white or grey literature), contribution \textit{type} (i.e.,~journal/conference paper for white literature, or book/blog post for grey literature), and \textit{year} of publication.}
\label{tab:selected-studies}
\end{table}

\section{A taxonomy for design principles, architectural smells and refactorings}
\label{sec:taxonomy}
\begin{figure*}
\centering
\includegraphics[width=.95\textwidth]{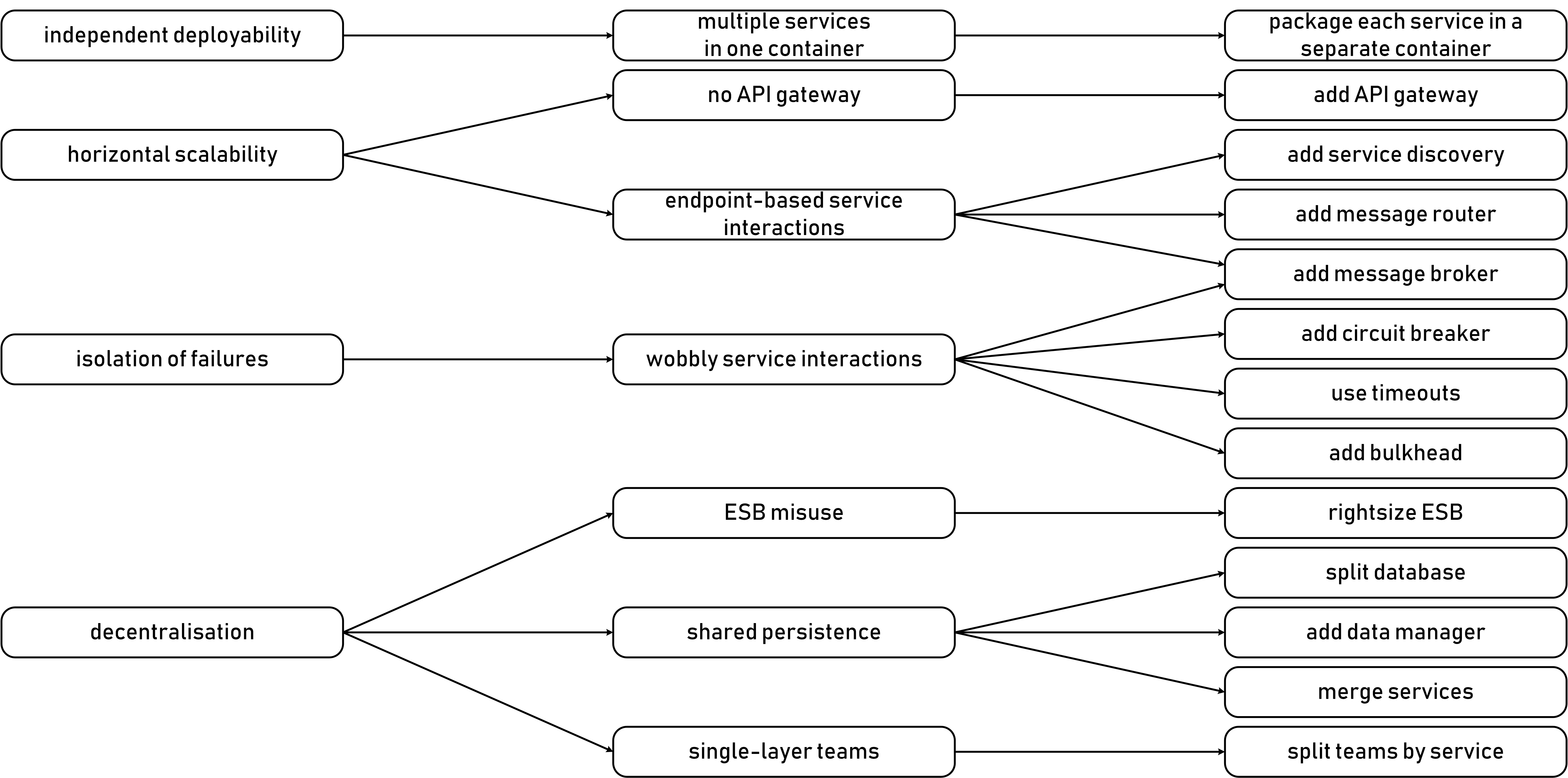}
\\
\hspace*{2cm} (a) \hfill (b) \hfill (c) \hspace*{2cm}
\caption{A taxonomy for (a) the design principles pertaining to the process viewpoint, (b) the architectural smells possibly violating such principles, and (c) the refactorings resolving such smells.}
\label{tab:taxonomy-antipatterns-refactorings}
\label{fig:taxonomy-antipatterns-refactorings}
\end{figure*}

\js{Fig.}~\ref{tab:taxonomy-antipatterns-refactorings} illustrates a taxonomy for the architectural smells pertaining to the considered design principles, and for the refactorings\footnote{For the sake of clarity, in the taxonomy we follow the naming of integration patterns proposed by Hohpe and Woolf~\cite{eip}.} allowing to \js{resolve} such smells.
We obtained our taxonomy by following the guidelines for conducting systematic reviews in software engineering proposed by Petersen et al.~\cite{petersen-systematic-reviews}:
\begin{enumerate}
    \item We established the \textit{design principles}, by aligning them with those pertaining to the process viewpoint (as per~\cite{microservices-tenets-m6}).
    \item We identified the \textit{architectural smells} by performing a first scan of the selected studies.
    \item We excerpted the concrete \textit{refactorings} directly from the selected studies after additional scans.
\end{enumerate}
The identified design principles, architectural smells and refactorings were manually organised to obtain a taxonomy.
\oz{The taxonomy underwent} various iterations among the authors of this study, and it was submitted for validation to an external expert.
This resulted in some corrections and amendments to the first version of the taxonomy, which resulted in the taxonomy displayed in \js{Fig.}~\ref{tab:taxonomy-antipatterns-refactorings}.
% OZ: I like the table very much, but would have expected a few more (and other) findings... e.g. isolate failure might also be achieved by coupling APIs more loosely (see MAP) 
% OZ: maybe summarize each refactoring briefly, as we cannot expect reviewers to be familiar with Circuit Breaker, Message Broker, etc. (?)
% for inspiration only: https://refactoring.guru/refactoring/smells 

\section{Architectural smells and refactorings}
\label{sec:antipatterns-refactorings}
\label{sec:smells-refactorings}
\begin{figure}
    \centering
    \includegraphics[width=\columnwidth]{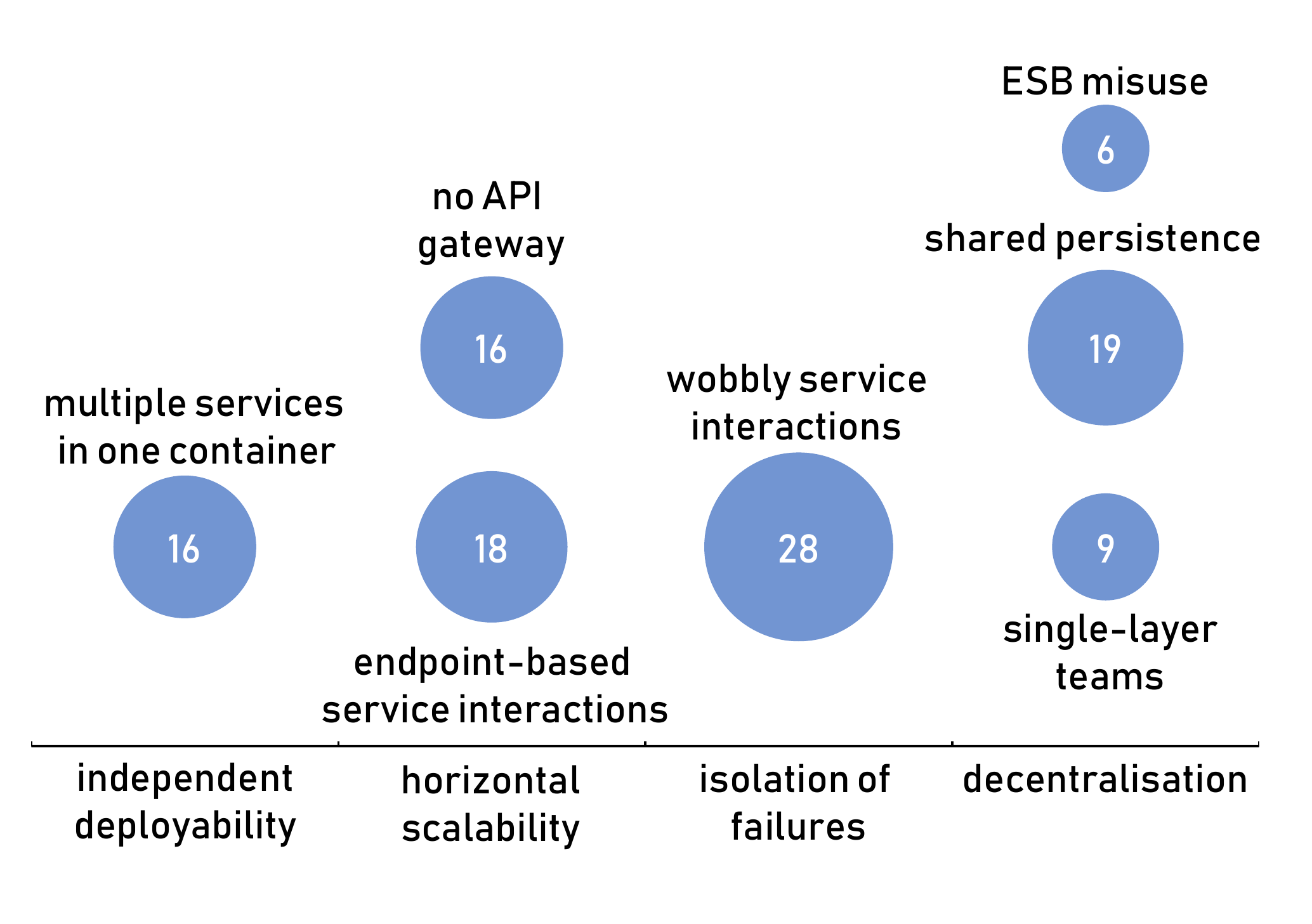}
    \caption{Coverage of the architectural smells in the selected studies. The size of each bubble is directly proportional to the number of selected studies discussing a refactoring pertaining to the corresponding smell. This number is also reported within each bubble.}
    \label{fig:antipatterns-coverage}
\end{figure}

%%%%%%%%%%%%%%%%%%%%%%%%%%%%%%%%%%%%%%%%%%%%%%%%%%%%%%%%%%%%%%%%%%%%%%%%%%%%%%%%%
% per ciascuna sottosezione, spiegare il principio e 
% illustrare ciascun antipattern con i rispettivi refactoring
%%%%%%%%%%%%%%%%%%%%%%%%%%%%%%%%%%%%%%%%%%%%%%%%%%%%%%%%%%%%%%%%%%%%%%%%%%%%%%%%%
Table~\ref{tab:antipatterns-coverage} shows the classification of all selected studies based on the taxonomy introduced in Sect.~\ref{sec:taxonomy}.
The table provides a first overview of the coverage of design principles, architectural smells and refactorings over the selected studies, despite (for reasons of readability and space) it only displays the classifications over the smells listed in the taxonomy\footnote{The detailed classification, displaying each occurrence of each refactoring, is publicly available at \url{https://github.com/di-unipi-socc/microservices-smells-and-refactorings}.}.
Such coverage is also displayed in Fig.~\ref{fig:antipatterns-coverage}, from which we can observe that all architectural smells in the taxonomy are significantly recognised by the authors of the selected studies, hence making it worthy to discuss them in detail.

%    \hline
%    & \textbf{independent}  & \multicolumn{2}{c}{\textbf{horizontal}}             & \multicolumn{2}{|c|}{\textbf{isolation}} & \multicolumn{3}{|c|}{\textbf{decentralisation}} \\
%    & \textbf{deployability} & \multicolumn{2}{c}{\textbf{scalability}}             & \multicolumn{2}{|c|}{\textbf{of failures}} & \multicolumn{3}{|c|}{ }                                              \\
%    & \textit{\js{multiple}} & \textit{\js{endpoint-based}} & \textit{no} &  &  & \textit{~} &  & \textit{single-} \\
%    & \textit{\js{services in}} & \textit{\js{service}} & \textit{API} & \textit{cascading} & \textit{no} & \textit{\js{ESB}} & \textit{shared} & \textit{layer}  \\
%    & \textit{\js{one container}} & \textit{\js{interactions}} & \textit{gateway} & \textit{failures} & \textit{timeouts}                   & \textit{\js{misuse}}     & \textit{persistence}        &  \textit{teams}        \\

\begin{table*}
\centering
\begin{tabular}{|c|c|c|c|c|c|c|c|}
    \hline
    ~
        & \centering \textbf{independent}  
        & \multicolumn{2}{c|}{\textbf{horizontal}}             
        & \centering \textbf{isolation} 
        & \multicolumn{3}{c|}{\textbf{~}} \\
        & \centering \textbf{deployab.}  
        & \multicolumn{2}{c|}{\textbf{scalability}}             
        & \centering \textbf{of failures} 
        & \multicolumn{3}{c|}{\textbf{decentralisation}} \\
    ~
        & \centering \textit{\js{multiple ser.}}
        & \centering \textit{~~~~no API~~~~} 
        & \centering \textit{\js{endpoint-bas.}}
        & \centering \textit{\js{wobbly}}
        & \centering \textit{\js{ESB}}
        & \centering \textit{~~~~shared~~~~}
        & \textit{~single-layer~} \\
        & \centering \textit{\js{in one cont.}}
        & \centering \textit{gateway} 
        & \centering \textit{\js{~~ser. inter.~~}}
        & \centering \textit{\js{~~ser. inter.~~}}
        & \centering \textit{\js{~~~~misuse~~~~}}
        & \centering \textit{~persistence~}
        & \textit{teams} \\
    \hline
    \cite{o4-seven-microservices-antipatterns}  
        &
        & $ \checkmark $ 
        &  
        &
        &
        &
        & \\
    \hline
    \cite{m19-systematic-mapping-study-microservice-architecture} 
        &
        &
        & $ \checkmark $
        & \js{$ \checkmark $}  
        &
        &
        & \\
    \hline
    \cite{m23-microservices-devops} 
        & $ \checkmark $       
        & 
        & $ \checkmark $
        & \js{$ \checkmark $}
        &
        &
        & $ \checkmark $  \\
    \hline
    \cite{m25-microservices-migration-patterns} 
        & $ \checkmark $       
        & $ \checkmark $   
        & $ \checkmark $  
        & \js{$ \checkmark $} 
        &
        & 
        & \\
    \hline
    \cite{o10-design-patterns-microservices} 
        &
        & $ \checkmark $   
        & $ \checkmark $  
        & \js{$ \checkmark $}  
        &
        & $ \checkmark $  
        & \\
    \hline
    \cite{b9-reactive-microservices} 
        &
        & $ \checkmark $ 
        & $ \checkmark $ 
        & \js{$ \checkmark $} 
        & $ \checkmark $ 
        & 
        & \\
    \hline
    \cite{b5-microservices-from-day-one}  
        &
        &
        &
        &
        &
        & $ \checkmark $  
        & \\
    \hline
    \cite{b11-spring-microservices} 
        & $\checkmark$ 
        & $\checkmark$ 
        & 
        & \js{$\checkmark$} 
        & 
        & $\checkmark$ 
        & \\
    \hline
    \cite{migrating-towards-microservices-m13} 
        & $ \checkmark $       
        &                            
        &                           
        &                           
        &                           
        &                           
        & $ \checkmark $  \\
    \hline
    \cite{o7-performance-patterns-microservices}  
        &
        &
        &
        & \js{$ \checkmark $}  
        &
        &
        & \\
    \hline
    \cite{m24-architecting-microservices} 
        &
        & $ \checkmark $   
        & $ \checkmark $  
        & \js{$ \checkmark $} 
        &
        &
        & \\
    \hline 
    \cite{research-on-microservices-m17} 
        &
        & $ \checkmark $   
        & $ \checkmark $  
        & \js{$ \checkmark $} 
        &
        &
        & \\
    \hline
    \cite{microservices-yesterday-today-tomorrow-m9}  
        & $ \checkmark $
        &
        &
        &
        &
        &
        & \\
    \hline
    \cite{migrating-enterprise-source-code-to-microservices-m8}  
        &
        &
        &
        &
        &
        & $ \checkmark $  
        & \\
    \hline
    \cite{o6-develop-great-microservices}  
        &
        &
        &
        &
        &
        &
        & $ \checkmark $  \\
    \hline
    \cite{o13-fundamentals-microservice-design} 
        & 
        & 
        & 
        & \js{$ \checkmark $} 
        &
        &
        & \\
    \hline
    \cite{o11-creating-microservice} 
        &
        &
        &
        &
        &
        & $ \checkmark $  
        & $ \checkmark $  \\
    \hline
    \cite{o8-microservices-in-practice}  
        & $ \checkmark $
        & 
        & $ \checkmark $   
        & \js{$ \checkmark $}  
        & $ \checkmark $  
        & $ \checkmark $  
        & \\
    \hline
    \cite{b6-microservices-for-enterprise}  
        & $ \checkmark $
        & $ \checkmark $ 
        & $ \checkmark $
        & \js{$ \checkmark $} 
        & $ \checkmark $  
        & $ \checkmark $ 
        & \\
    \hline
    \cite{microservices-journey-challenges-m7}  
        & $ \checkmark $       
        &
        &
        & \js{$ \checkmark $} 
        &
        &
        & \\
    \hline
    \cite{m27-challenges-monolith-microservices} 
        &
        &
        &
        & \js{$ \checkmark $}  
        &
        & $ \checkmark $  
        & $ \checkmark $  \\
    \hline
    \cite{m11-microservices-for-legacy-software}  
        &
        &
        &
        & \js{$ \checkmark $}  
        &
        & $ \checkmark $  
        & \\
    \hline
    \cite{b10-microservices-patterns-applications} 
        & $ \checkmark $ 
        & $ \checkmark $ 
        & $ \checkmark $ 
        & \js{$ \checkmark $} 
        & 
        & 
        & \\
    \hline
    \cite{o2-microservices-fowler}  
        &
        & 
        & $ \checkmark $
        & \js{$ \checkmark $}  
        & $ \checkmark $  
        & 
        & $ \checkmark $  \\
    \hline
    \cite{o12-container-patterns} 
        & $ \checkmark $
        &
        &
        &
        &
        &
        & \\
    \hline
    \cite{o9-power-patterns-microservices}  
        &
        &
        & $ \checkmark $
        & \js{$ \checkmark $}  
        &
        &
        & \\
    \hline
    \cite{b4-microservice-architecture}  
        &
        & $ \checkmark $   
        & $ \checkmark $  
        & \js{$ \checkmark $}  
        &
        & $ \checkmark $  
        & $ \checkmark $  \\
    \hline
    \cite{b1-building-microservices}  
        & $ \checkmark $       
        &
        & $ \checkmark $   
        & \js{$ \checkmark $}
        &
        &
        & \\
    \hline
    \cite{b7-release-it}  
        & $ \checkmark $ 
        & $ \checkmark $ 
        & $ \checkmark $ 
        & \js{$ \checkmark $}  
        & 
        & 
        & \\
    \hline
    \cite{b3-microservices-anti-patterns-pitfalls}  
        &
        &
        &                           
        & \js{$ \checkmark $}  
        &
        & $ \checkmark $ 
        & \\
    \hline
    \cite{o1-microservices-intro}  
        &
        & $ \checkmark $  
        & $ \checkmark $  
        &
        &
        & $ \checkmark $  
        & \\
    \hline
    \cite{o3-microservices-io}  
        &
        & $ \checkmark $ 
        &  
        & \js{$ \checkmark $}
        &
        & $ \checkmark $  
        & \\
    \hline
    \cite{o14-common-pitfalls}  
        & 
        & 
        & 
        & \js{$ \checkmark $} 
        & 
        &
        & \\
    \hline
    \cite{o5-microservices-antipatterns}  
        &
        &
        & $ \checkmark $   
        & \js{$ \checkmark $}
        &
        & $ \checkmark $  
        & \\
    \hline
    \cite{microservices-mjolnirr-case-study-m14} 
        & $ \checkmark $
        &
        &
        &
        &
        &
        & \\
    \hline
    \cite{m26-microservices-pains-gains} 
        & $ \checkmark $
        & $ \checkmark $  
        & 
        & \js{$ \checkmark $} 
        &
        & $ \checkmark $  
        & \\
    \hline
    \cite{microservice-bad-smells-m5}  
        &
        & $ \checkmark $  
        & $ \checkmark $  
        &
        & $ \checkmark $  
        & $ \checkmark $  
        & \\
    \hline
    \cite{processes-motivations-issues-microservices-m16} 
        &
        &
        &                           
        &
        &
        & $ \checkmark $  
        & $ \checkmark $  \\
    \hline
    \cite{architectural-patterns-microservices-m3} 
        & $ \checkmark $       
        &
        &
        &
        &
        & $ \checkmark $  
        & \\
    \hline
    \cite{b8-microservices-flexible-sw-architecture} 
        & 
        &
        & $ \checkmark $ 
        & \js{$ \checkmark $} 
        & 
        & $ \checkmark $ 
        & $ \checkmark $ \\
    \hline
    \cite{microservices-tenets-m6}  
        & $ \checkmark $
        &                            
        &
        &
        & $ \checkmark $  
        &
        & \\
    \hline
\end{tabular}
\caption{Classification of the selected studies based on \ab{the taxonomy \js{in Fig.~\ref{tab:taxonomy-antipatterns-refactorings}}}.
%for design principles, architectural smells and refactorings for microservices.
}
\label{tab:antipatterns-coverage}
\end{table*}

We hereafter illustrate how (according to the authors of the selected studies) each design principle can be affected by each corresponding architectural smell, as well as how each smell can be \js{resolved} by applying a corresponding refactoring. 
When multiple refactorings are applicable to \js{resolve} an architectural smell, to provide a first measurement of how much a refactoring is used to \js{resolve} it, we display the weight\footnote{We \ab{measure} the weight of a refactoring as the percentage of its occurrences among all occurrences of all refactorings for the same smell. This is analogous to what done by Pahl et al.~\cite{cloud-container-technologies} to measure weights while classifying studies on cloud container technologies.} of each refactoring by exploiting \%-based pie charts.

\subsection{Independent deployability}
% Newman: independent deployability
% FowlerLewis: independent deployability
In microservice-based applications, each microservice should be operationally independent from the others, meaning that it should be possible to deploy and undeploy a microservice indepedently from the others~\cite{b1-building-microservices}.
This indeed impacts on the initial deployment of a microservice, which can get started without waiting for other microservices to be running, as well as on the possibility of adding/removing replicas of a microservice at runtime.

We discuss below the \js{\textsc{Multiple Services in One Container}} smell, showing how it violates the above principle and how it can be \js{resolved}.

%\subsubsection{multiple services in one container}
\smallskip \noindent
%{\bf multiple services in one container.}
{\bf \ab{Multiple services in one container}.}
Containers (such as Docker container\ab{s}) provide an ideal way to deploy microservices addressing the above requiremen\ab{t}, if properly used.
Each microservice can indeed be packaged in a container image, and different instances of a same microservice can be launched by spawning different containers from the corresponding image.
With this view, the orchestration of the deployment and management of a microservice-based application can be performed by exploiting the currently available support for orchestrating Docker containers~\cite{o8-microservices-in-practice}.

The above is the right way of using containers, at least according to the authors of 16 of the selected studies. 
They indeed highlight how placing multiple services in one container would constitute an architectural smell for the independent deployability of microservices.
If two microservices would be packaged in the same Docker image, spawning a container from such image would result in launching both microservices. 
Similarly, stopping the container would result in stopping both microservices.
In other words, by placing two microservices in the same container, \ab{these services} would operationally depend one another, as it would not be possible to launch a new instance of one of such microservices, without also launching an instance of the other.

If the \js{\textsc{Multiple Services in One Container}} smell occurs, the solution is to refactor the application in such a way that %containers are used as discussed above, i.e.,~
\ab{each} microservice is packaged in a separate container image.

\subsection{Horizontal scalability}
% Newman: (conseguenza di) independent deployability
% FowlerLewis: (conseguenza di) componentisation via services
The possibility of adding/removing replicas of a microservice is a direct consequence of the independent deployability of microservices.
To ensure its horizontal scalability, all the replicas of a microservice \js{$m$} should be reachable by \js{the microservices invoking $m$}~\cite{o8-microservices-in-practice}. 

In the selected studies, two architectural smells em\-erged as possibly violating the horizontal scalability of microservices, i.e.,~\js{\textsc{Endpoint-based Service Interactions}} and \textsc{No API Gateway}, which we discuss hereafter.

%\subsubsection{direct service interactions}
\smallskip \noindent 
{\bf \js{Endpoint-based service interactions}.}
% oz3: the new small caps font seems to cause lines to run into margin?
% oz3: spell out numbers up to "twelve", while 13 etc. stay as is
%The \js{\textsc{Endpoint-based Service Interactions}} 
\js{This} smell occurs in an application when one or more of its microservices invoke a specific instance of another microservice (e.g.,~because its location is hardcoded in the source code of the microservices invoking it, or because no load balancer is used).
If this is the case, when scaling out the latter microservice by adding new replicas,
\ab{these} cannot be reached by the invokers, hence only resulting in a waste of resources. %\footnote{For instance, consider a microservice $m_1$ directly invoking a microservice $m_2$, and suppose that after adding two new instances of $m_2$ (say, $m_2'$ and $m_2''$), $m_1$ still keeps invoking only $m_2$.
%The resources used to run $m_2'$ and $m_2''$ are hence wasted, as such instances are never used.}.

\begin{figure}
    \centering
    \includegraphics[width=\columnwidth]{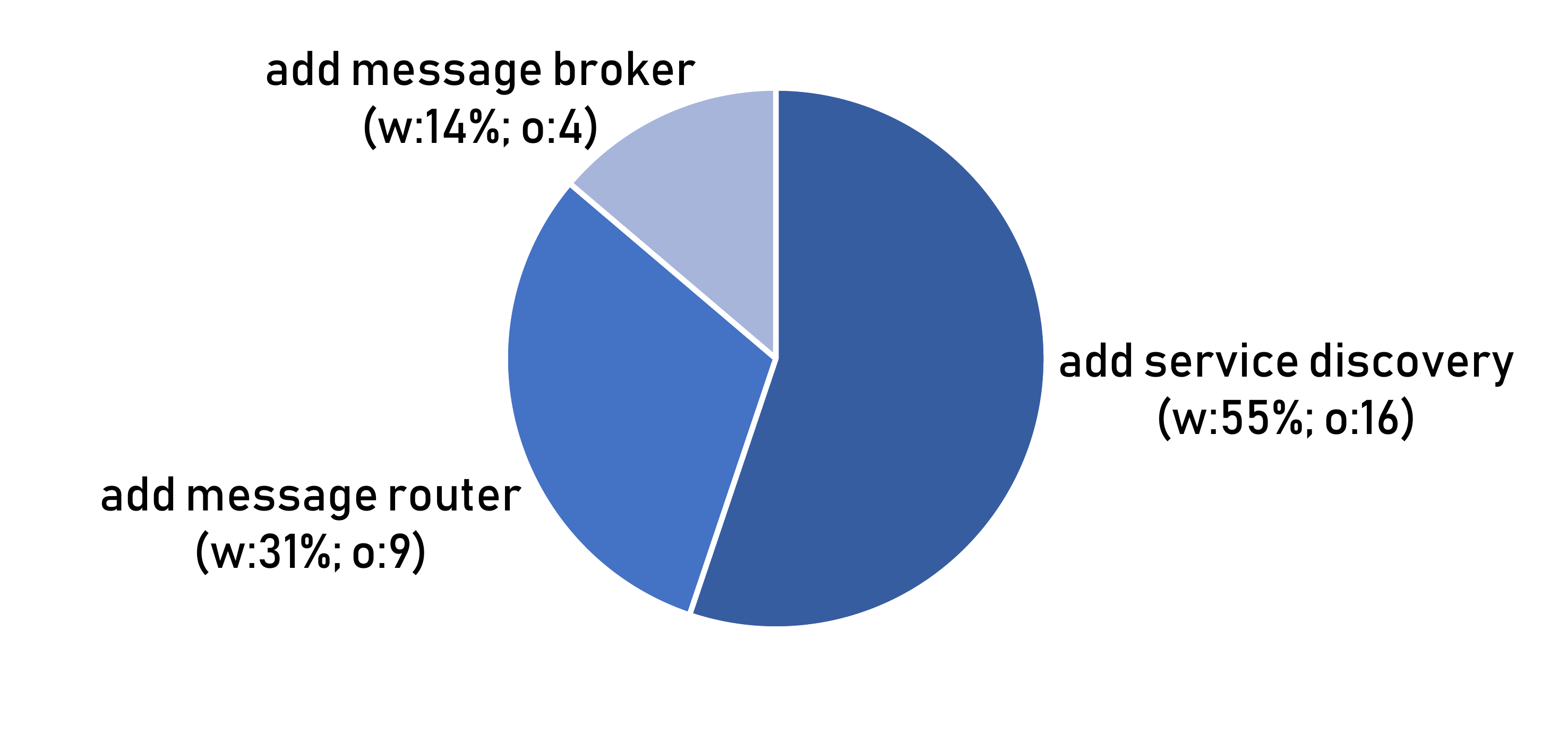}
    \caption{Weights (\textsf{w}) and occurrences (\textsf{o}) of the refactorings for the \js{\textsc{Endpoint-based Service Interactions}} smell.}
    \label{fig:refactoring-direct-service-interaction}
\end{figure}

From the selected studies, \abb{it became evident that the} \js{\textsc{Endpoint-based Service Interactions}} smell can be \js{resolv}ed by applying three different refactorings (Fig.~\ref{fig:refactoring-direct-service-interaction}). 
The most common solution is to introduce a service discovery mechanism.
\ozz{Such mechanism can be implemented} as a service storing the actual locations of all instances of the microservices in an application~\cite{o3-microservices-io}.
Microservice instances send their locations to the service registry \ab{at} startup, and they are unregistered \ab{at}  shutdown.
When wanting to interact with a microservice, a client can then query the service discovery to retrieve the location of one of its instances.

The other two possible solutions share the same \ab{goal}, i.e.,~decoupling the interaction between two microservices by introducing an intermediate integration pattern.
\js{Nine} of the selected studies indeed suggest to introduce a message router (e.g.,~a load balancer), so that the requests to a microservices are routed towards all its actual instances. \js{Four} of the selected studies instead suggest to exploit message brokers (e.g.,~message queues) to decouple the interactions between two or more microservices.

%\subsubsection{No API gateway}
\smallskip \noindent 
{\bf No API gateway.}
When a microservice-based application lacks an API gateway, the clients of the application necessarily have to invoke its microservices directly.
The result is a situation similar to that of the \js{\textsc{Endpoint-based Service Interactions}} smell, with the invoker being a client of the application.
The client indeed interacts only with the specific instances of the microservices it needs.
If one of such microservices is scaled out and the client still keeps invoking the same instance of the microservice, then we have a waste of \abb{resources}.
%resources (as we need additional resources to run the new instances of the scaled microservices, which are never used).

The authors of all the selected studies discussing the \js{\textsc{No API Gateway}} smell agree that the solution to this smell is to add 
\abb{one} API gateway
%one or more API gateways 
to the application. 
The latter act as single entry points for all clients, and they handle requests either by routing them or by fanning them out to the instances of the microservices that must handle them~\cite{o3-microservices-io}.

It is worth noting that, even if the \js{\textsc{No API Gateway}} smell results in a similar situation to that of the \js{\textsc{Endpoint-based Service Interactions}} smell, the refactorings to \js{resolv}e them are different.
The reason for this resides in the main difference between the two architectural smells.
\abb{The} \textsc{No API Gateway} smell occurs at the edge of the architecture of a microservice-based application, with the clients of the application directly invoking its microservices, while the \js{\textsc{Endpoint-based Service Interactions}} smell occurs in between its microservices~\cite{b4-microservice-architecture}.
Given this, the introduction of an API gateway can be useful not only for facilitating the horizontal scalability of the microservices forming an application, but also for various other reasons.
For instance, rather than implementing end-user authentication or throttling 
%\abb{or} orchestration 
in each microservice, these can be implemented once for the whole application in the API gateway~\cite{o4-seven-microservices-antipatterns}.
% oz (0315): pls double check: is "orchestration" the right term here? meaning clustering (Kubernetes style) or composition (BPEL-like)? overloading an API Gateway is seen as a smell by ThoughtWorkers, btw

\subsection{Isolation of failures}
% Newman: isolated failure
% FowlerLewis: design for failure
Microservices can fail for many reasons (e.g.,~network or hardware issues, application-level issues, bugs), hence becoming unavailable to serve other microservices.
Additionally, communication fails from time to time in any kind of distributed system, and this is even more likely to occur in microservice-based systems, simply because of the amount of messages \abb{exchanged} among microservices~\cite{microservices-journey-challenges-m7}.
Microservice-based applications should hence be designed so that each microservice can tolerate the failure of any invocation to the microservices it depends on~\cite{o2-microservices-fowler}.
If this is ensured, then a microservice-based application results to be much more resilient than a monolithic application, simply because failures affects only few microservices in an application, instead of the whole monolith~\cite{b1-building-microservices}.

The authors of the selected studies identify and discuss \js{an architectural smell} that can \js{possibly} violate the isolation of failures in microservice-based solutions.
%These are the \js{\textsc{Cascading Failures}} smell and the \js{\textsc{No Timeouts}} smell,
\js{This is the \textsc{Wobbly Service Interactions} smell}, which we discuss hereafter.

\smallskip \noindent
{\bf Wobbly service interactions.}
\js{The interaction of a microservice $m_i$ with another microservice $m_f$ is ``wobbly" when a failure in $m_f$ can result in triggering a failure also in $m_i$. 
This typically happens when $m_i$ is directly consuming one or more functionalities offered by $m_f$, and $m_i$ is not provided with any solution for handling the possibility of $m_f$ to fail and be unresponsive. 
If this is the case, $m_i$ will also fail in cascade, and (in a worst case scenario) the failure of $m_i$ can result in triggering the failure of other microservices, which in turn trigger other cascading failures, and so on~\cite{microservices-journey-challenges-m7}.}

\begin{figure}
    \centering
    \includegraphics[width=\columnwidth]{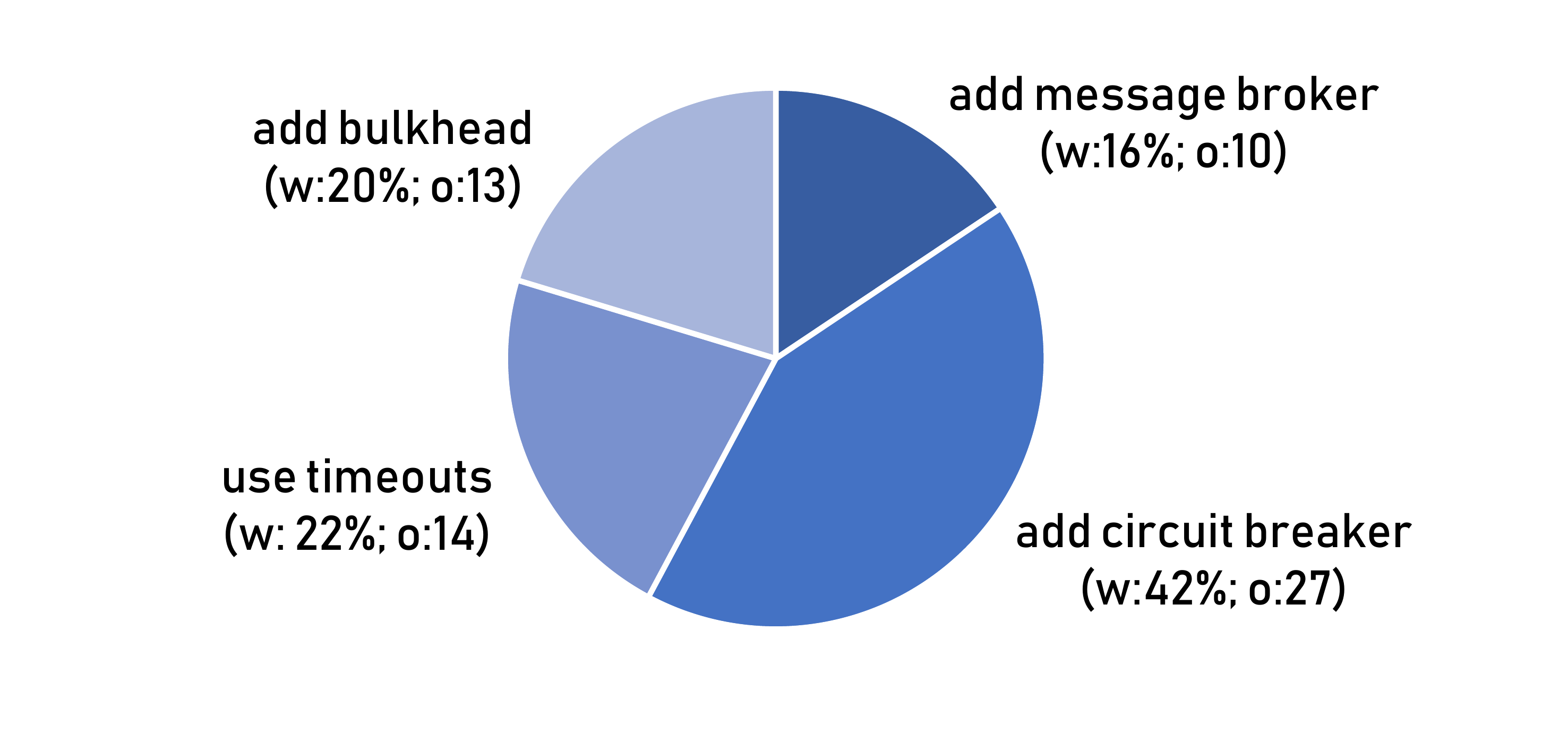}
    \caption{\js{Weights (\textsf{w}) and occurrences (\textsf{o}) of the refactorings for the \js{\textsc{Wobbly Service Interactions}} smell.}}
    \label{fig:refactoring-wobbly-service-interaction}
\end{figure}

\js{To avoid \textsc{Wobbly Service Interactions} (such as the one between $m_i$ and $m_f$ described above), the authors of the selected studies identify four possible solutions (Fig.~\ref{fig:refactoring-wobbly-service-interaction}).}
%
%\subsubsection{Cascading failures}
%\smallskip \noindent 
%{\bf Cascading failures.}
%A cascading failure occurs when the failure of one microservice triggers the failures of other microservices, which directly or indirectly depend on the failed microservice~\cite{microservices-journey-challenges-m7}.
%An obvious example is a microservice $m_i$ directly invoking another microservice $m_f$, with the latter failing.
%If the microservice $m_i$ is not provided with any solution for handling the possibility of $m_f$ to fail, then $m_i$ will also start to fail, resulting in a cascading failure.
%In a worst-case scenario, the failure of $m_i$ triggers the failure of other microservices, which in turn trigger other cascading failures, and so on.
%As a result, we may have that an entire portion of the application \ab{fails} in cascade to the failure of a single microservice simply because of the initial failure of $m_f$.
%
%To avoid \ab{that}, the authors of the selected studies identify three possible solutions (Fig.~\ref{fig:refactoring-cascading-failures}).
%
%\begin{figure}
%    \centering
%    \includegraphics[width=\columnwidth]{dat/refactoring-cascading-failures.pdf}
%    \caption{Weights (\textsf{w}) and occurrences (\textsf{o}) of the refactorings for the \js{\textsc{Cascading Failures}} smell.}
%    \label{fig:refactoring-cascading-failures}
%\end{figure}
%
The most common solution is the usage of a circuit breaker to wrap the invocations from a microservice to another.
In the normal ``closed" state, the circuit breaker forwards the invocations to the wrapped microservice, and it monitors their execution to detect and count failing invocations.
Once the frequency of failures \ab{reaches} a certain (customisable) threshold, the circuit breaker trips and ``opens" the circuit.
All further calls to the wrapped microservice will ``safely fail", as the circuit breaker will immediately return an error message to the calling microservices.
The latter can then exploit the error messages returned by the circuit breaker to avoid failing themselves~\cite{o2-microservices-fowler}.

Following the same baseline idea of circuit breakers, \js{ten} of the selected studies propose to decouple the interaction between invoking and invoked microservices by exploiting a message broker (e.g.,~a message queue).
The usage of a broker \ab{allows} the invoker to send its requests to the broker, and \ab{allows} the invoked microservice to process such requests when it is available.
In this way, there is no direct interaction between the two microservices, and the invoker does not fail when the invoked microservice fails (as the former continues to send messages to the broker).
\abb{On the other hand,}
%At the same time, 
the usage of message brokers is more costly \ozz{compared to} circuit breakers. 
The reason is that message brokers require to intervene on the interaction protocol between two microservices, which should start putting and getting messages to/from the broker.
Instead, with circuit breakers the interaction protocol between two microservices is \ab{unaltered}, as a circuit breaker simply wraps the invocation of a microservice.
This is the reason why message brokers are much less discussed \ozz{than} circuit breakers.

\js{The most discussed alternative to circuit breakers are however timeouts, which are a simple yet effective mechanism allowing a microservice to stop waiting for an answer from another microservice, when the latter is unresponsive (e.g.,~since it failed or due to network issues).
Well-placed timeouts provide fault isolation, as the fact that a microservice is unresponsive does not create any other issue in the microservices invoking it~\cite{b1-building-microservices}.
However, such a kind of solution \ozz{might not likely} to be applicable nowadays, as \oz{some of} the APIs used to remotely invoke microservices have few or no explicit timeout settings~\cite{b1-building-microservices}.} 
% OZ (0315): I agree with DN. So weakened statement, HTTP afaik has timeouts
\oz{Note that the timeout can be also set in the invoker (e.g., by setting the timeout on an HTTP request), hence it is not always requested to have a timeout setting on the invoked service.}

\js{Finally, another alternative} is the usage of bulkheads, whose ultimate goal is to enforce the principle of damage containments \js{(like bulkheads in ships, which prevent water to flow across sections)}.
The idea is that, if cascading failures cannot be avoided, they should at least be limited by exploiting bulkheads.
More precisely, the microservices forming an application should be \ozz{logically and/or physically} partitioned so as to ensure that the failure of a microservice can be propagated at most to the other microservices in the same partition, by preventing the rest of the system from being affected by such failure~\cite{b7-release-it}.

\subsection{Decentralisation}
% Newman: decentralise everything
% FowlerLewis: decentralise data management, smart endpoints & dumb pipes, decentralise governance
Decentralisation should occur in all aspects of micro\-service-based applications~\cite{b1-building-microservices}.
This also means the business logic of an application should be fully decentralised and distributed among its microservices, each of which should own its own domain logic~\cite{microservices-tenets-m6}.

The authors of the selected studies indentify and discuss three architectural smells possibly violating the above principle, i.e.,~the \js{\textsc{ESB Misuse}, \textsc{Shared Persistence} and \textsc{Single-layer Teams}} smells.
We hereafter discuss them, by also illustrating the refactorings currently employed to \js{resolv}e them.

% OZ (round 2): rewritten, please review and check for consistency with other sections
%\subsubsection{ESB misplacement and/or abuse}
\smallskip \noindent 
{\bf \js{ESB misuse}.}
% OZ: This one, unlike the others that I agree with, might need some work and discussion. Mostly a wording and positioning issue.
% OZ: Some of the "anti-ESB climate" in the microservices community has its root cause in vendor tactics an poor implementations, the pattern itself cannot be blamed. For instance, ESBs do not have to be centralistic. And the "smart endpoints, dumb pipes" recommendation actually goes back to ESB days (I have a reference in the tenets paper to provide evidence for this statement). 
% OZ: btw, many ESB products usually are queue-based message brokers :-)
% OZ: we might want to revisit the reference that talks about ESB vs. Message Broker as alternatives 
The misuse of Enterprise Service Buses (ESB) products is considered to be an architectural smell by the microservice community. When positioned as a single central hub (with the services as spokes), an ESB may become a bottleneck both architecturally and organizationally~\cite{Pautasso:2017:MSIP}. 
``Smart endpoints \& dumb pipes" has been a recommended practice since the very beginnings of service-oriented architectures~\cite{microservices-tenets-m6} that regrettably has not always been followed in all SOA implementations. Such ESB abuse may lead to undesired centralisation of business logic and dumb services~\cite{b1-building-microservices}. The microservices community therefore (re-)emphasizes the  decoupling of microservices and their cohesiveness~\cite{o2-microservices-fowler}.

Whenever a central ESB is used for connecting microservices in an application, the topology should be refactored to remove the dependency on a single middleware component instance. 
% In this perspective, all the selected studies discussing the ESB-based interaction smell recommend to replace ESBs with message brokers (e.g.,~message queues).
\js{Multiple instances should instead be used, and they should implement queue-based asynchronous messaging. The latter}
%Queue-based asynchronous messaging 
only permits adding and removing messages, hence forming  a ``dumb pipe". 
The ``smart" part \js{should be} left to the microservices, which implement the logic for deciding when/how to process the messages in the message broker~\cite{microservice-bad-smells-m5}.
Additional infrastructure logic, for instance traffic management capabilities, may be placed in side cars accompanying each service. 
This repositioning and rectification of ESB middleware improves the decoupling characteristics of the services architecture and reestablishes the original ``smart endpoints \& dumb pipes" recommendations from the first wave of service-orientation. 

%\subsubsection{Shared persistence}
\smallskip \noindent
{\bf Shared persistence.}
The \js{\textsc{Shared Persistence}} smell occurs whenever two microservices access and manage the same database\ab{, possibly violating  the decentralisation design principle~\cite{m26-microservices-pains-gains}}. 
%If this is the case, it naturally follows that they do not own their own domain logic, and the decentralisation design principle gets violated~\cite{m26-microservices-pains-gains}. 

\begin{figure}
    \centering
    \includegraphics[width=\columnwidth]{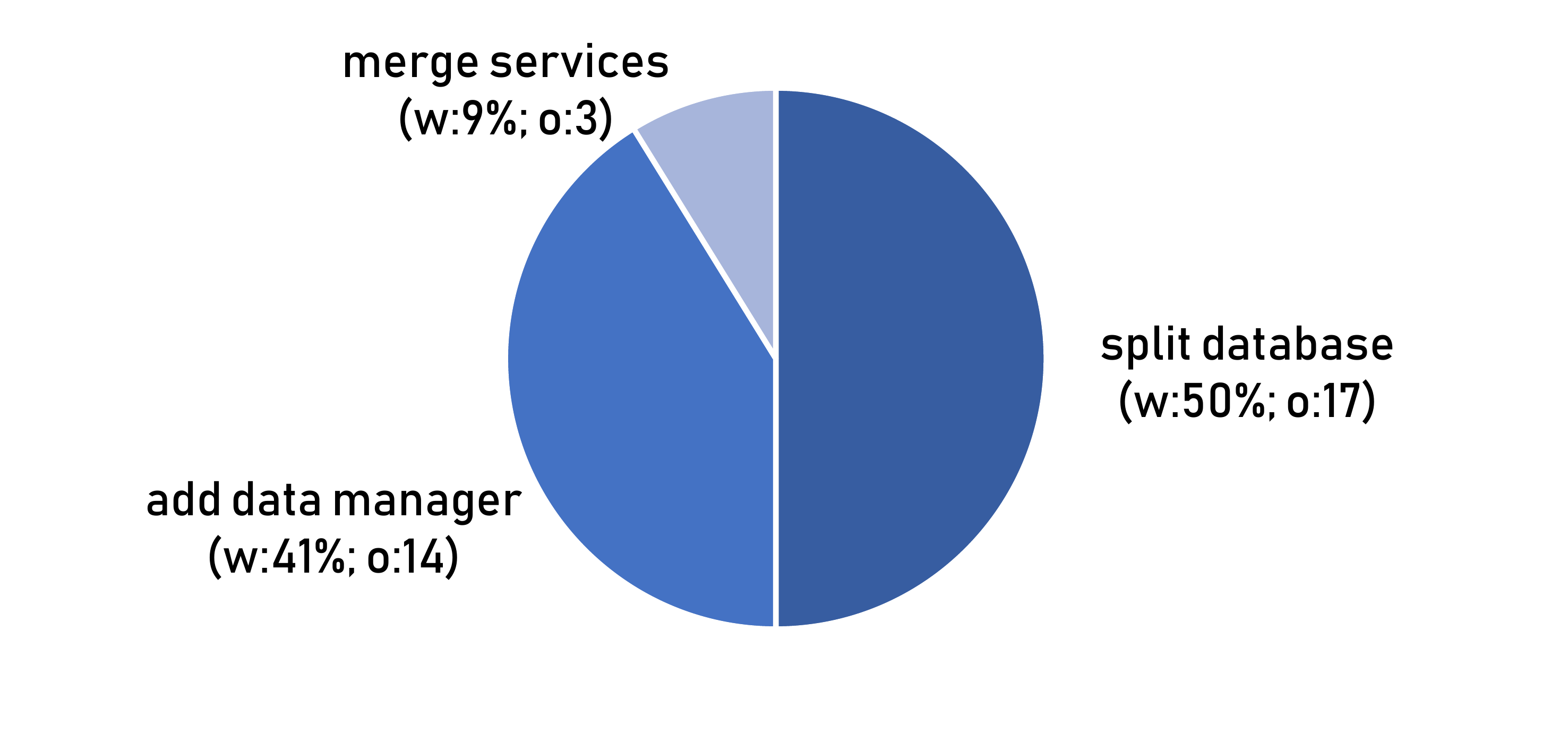}
    \caption{Weights (\textsf{w}) and occurrences (\textsf{o}) of the refactorings for the \js{\textsc{Shared Persistence}} smell.}
    \label{fig:refactoring-shared-persistency}
\end{figure}

The three currently available solutions for refactoring microservices and \js{resolving} the \js{\textsc{Shared Persistence}} smell are shown in Fig.~\ref{fig:refactoring-shared-persistency}.

\ozz{Although the ultimate goal of these three solutions is the same (i.e.,~having each database accessed by only one microservice), they are very diverse in spirit. They apply to different situations, highly depending on the microservices accessing the same database.}

The most discussed solution is to actually split a database shared by multiple microservices, in such a way that each microservice \ab{accesses and manages} only the data it needs.
This solution is the one requiring less intervention on the microservices, as they would continue to use the same protocol to interact with the databases. 
At the same time, splitting a database into a set of independent databases is not always possible or easy to achieve.
Also, if some data is to be replicated among the databases obtained from the split, then mechanisms for \ab{(eventual)} data consistency should be introduced after the refactoring~\cite{m26-microservices-pains-gains}.
Given the above, the split of database is recommended when the microservices accessing the same database implement separate business logics working on disjoint portions of such database~\cite{b6-microservices-for-enterprise}.
% oz (0315): parked for later consideration: in our Microservices API Pattern language, we call the data managers *Information Holders* (a tribute to RDD): https://preview.microservice-api-patterns.org/patterns/responsibility/

\ab{The}  most discussed alternative is to introduce an additional microservice, acting as ``data manager".
The data manager becomes the only microservice interacting \ab{with} and managing  the database, and the microservices that were accessing the database now have to interact with the data manager to ask for accessing and updating the data.
While this solution introduces some additional communication overhead, it is considered as always applicable, and the data manager can also be enriched with additional logic for processing the data it manages~\cite{b6-microservices-for-enterprise}.

Finally, it is worth commenting on the refactoring discussed in three of the selected studies, i.e.,~merging the microservices accessing the same database.
The idea is that, when multiple microservices access the same database, this \ab{may} be a signal of the fact that the application has been split too much, by obtaining too fine-grained microservices processing the same data.
If this is the case, then the possibility of merging such microservices is a concrete option to be evaluated~\cite{microservice-bad-smells-m5}.

%\subsubsection{Single-layer teams}
\smallskip \noindent
{\bf Single-layer teams.}
To maximize the autonomy that microservices make possible, the governance of microservices should be decentralised and delegated to the teams that own the microservices themselves. 
\ozz{As pointed out by Zimmermann~\cite{microservices-tenets-m6}, even if this is not a technical concern, it is related to the process viewpoint due to its cross-cutting nature.}
\js{The microservice community indeed strongly emphasizes the connection between architecture and organisation, especially concerning the integration of the microservices in an application~\cite{o6-develop-great-microservices,o11-creating-microservice,o2-microservices-fowler}}.

The classical approach of splitting teams by technology layers (e.g.,~user interface teams, and middleware teams, and database teams) is hence considered an architectural smell, as \oz{any change to} a microservice \js{may result} in a cross-team project having take time and budgetary approval~\cite{o2-microservices-fowler}.
% oz /0315): reworded as per DBs comment:
\oz{This may be the case for the refactorings  discussed so far.} %, in this section, }
% each of which \js{could} require cross-team interactions.

The microservice approach to team splitting is orthogonal to the above, as each microservice should be assigned to a full-stack team whose members span across all technology layers.
In this way, the interactions for updating a microservice (e.g.,~to apply one of the refactorings discussed in this section) are limited to the team managing such microservice, which can independently decide how to proceed and implement the updates~\cite{migrating-towards-microservices-m13}.

In short, if the governance of a microservice-based is organised by \js{\textsc{Single-layer Teams}}, this is an architectural smell.
The solution is to split teams by microservice, \ozz{rather} than by technology layer~\cite{o2-microservices-fowler}. 

\section{Threats to validity}
\label{sec:ttv}
Following the taxonomy developed by Wohlin et al.~\cite{wohlin}, four potential theats may affect the validity of our study. 
These are the threats to \textit{external} validity, the threats to \textit{internal} and \textit{construct} validity, and the threats to \textit{conclusions} validity, which we discuss hereafter.

%\subsection{Threats to external validity}
\smallskip \noindent
{\bf External validity.}
As per Wohlin et al.~\cite{wohlin}, the external validity concerns the applicability of a set of results in a more general context.
Since we selected the primary studies from a very large extent of online sources, the identified architectural smells and refactorings may only be partly applicable to the broad area of disciplines and practices on microservices, hence threatening external validity.

To reinforce the external validity of our findings, we organised \js{two} feedback sessions during our analysis of the existing literature.
% feedback sessions: microservices @ foclasa, davide phd discussion
We analysed the discussion following-up from the feedback session, and we exploited this qualitative data to fine-tune both our research methods and the applicability of our findings.
We also prepared a GitHub repository\footnote{\url{http://github.com/di-unipi-socc/microservices-smells-and-refactorings}.}, where we placed the artifacts produced during our analysis, so as to make it available to all who wish to \ab{deepen} their understanding on the data we produced.
We believe that this can help in making our results and observations more explicit and applicable in practice.

Additionally, one may argue that our selection criteria are too restrictive.
The rationale behind such criteria is that we aim focusing only on representative studies, by requiring selected studies to discuss at least an architectural smell \abb{{\em and}} a refactoring for \js{resolv}ing it.
There is however a risk of having missed some relevant literature, as a study might not explicitly mention the architectural smells and refactorings in our taxonomy (\js{Fig.}~\ref{tab:taxonomy-antipatterns-refactorings}).
To mitigate this threat, we carefully checked both selection criteria against each candidate study, by verifying whether a study was discussing the problems characterised by an architectural smell, and whether it was discussing the architectural changes characterising a refactoring.
Even if a study was not explicitly referring to a smell/refactoring, but it was reporting on the corresponding problems/changes, the study was included in the selected literature.

Finally, there is a risk of having missed relevant grey literature, since industrial studies may exploit a different terminology than ours (e.g.,~a blog post discussing some architectural smells and refactorings may not employ the term ``smell" or ``refactor").
To mitigate this threat to validity, we included relevant synonyms in the search string, and we 
exploited the features offered by search engines, which naturally support including related terms in string-based searches.

%\subsection{Threats to construct and internal validity}
\smallskip \noindent
{\bf Construct and internal validity.}
The internal validity concerns the validity of the method employed to study and analyse data (e.g.,~the potential types of bias involved), while the construct validity concerns the generalisability of the constructs under study~\cite{wohlin}.

To mitigate the corresponding potential threats, the obtained taxonomy \ab{underwent} various iterations among the authors of this study to avoid bias by triangulation, and it was submitted for validation to an external expert. % (see Sect.~\ref{sec:taxonomy}).
The same process was applied to the classification of the selected studies, and to the results of the analysis.

%\subsection{Threats to conclusions validity}
\smallskip \noindent
{\bf Conclusions validity.}
The conclusions validity concerns the degree to which the conclusions of a study are reasonably based on the available data~\cite{wohlin}.

In this perspective, and with the aim of performing a sound analysis of the data we retrieved, we exploited inter-rater reliability assessment to limit potential biases in our observations and interpretations.
Additionally, the observations and conclusions discussed in this paper were independently drawn, and they were then double-checked against the selected studies and related studies in a joint discussion session. 

\section{Related work}
\label{sec:related}
There exist various studies on microservices, aimed at analysing and classifying the state of the art and practice on microservices.
% existing reviews on microservices (not discussing architectural smells - hence different from ours)
Pahl and Jamshidi~\cite{microservices-sms} and Taibi et al.~\cite{architectural-patterns-microservices-m3} present two first systematic mapping studies on microservices.
Pahl and Jamshidi~\cite{microservices-sms} discuss agreed and emerging concerns on microservices, position microservices with respect to current cloud and container technologies, and elicit potential research directions.
Taibi et al.~\cite{architectural-patterns-microservices-m3} instead report on architectural patterns common to microservice-based solutions, by discussing the advantages, disadvantages and lessons learned of each pattern.
However, neither Pahl and Jamshidi~\cite{microservices-sms} nor Taibi et al.~\cite{architectural-patterns-microservices-m3} provide an overview both on the architectural smells applicable to microservices and on the refactorings for \js{resolv}ing such smells.

Two other examples \ab{are} the industrial surveys by Di~Francesco et al.~\cite{migrating-towards-microservices-industrial-survey} and by Ghofrani and L\"ubke~\cite{challenges-microservices-industrial-survey}, which both discuss the current state of practice on microservices in the IT industry.
Both report on empirical studies conducted in the form of surveys for practictioners working everyday with microservices, to elicit the challenges and advantages on employing microservices.
This differs from our study, as we aim at distilling the architectural smells that can affect the architecture of a microservice-based solution, as well as the refactorings allowing to \js{resolv}e such smells.

%pains and gains
Similar considerations apply to the systematic review by Soldani et al.~\cite{m26-microservices-pains-gains}, who provide an overview on the state of practice on microservices.
Soldani et al. systematically analyse the grey literature on microservices, in order to identify the technical/operational advantages and disadvantages of the microservice-based architectural style.
The objective of Soldani et al. hence differs from ours, as we aim at discussing concrete architectural smells and refactorings for the microservice-based architectural style.

%existing work on smells/architectural smells microservices
In this perspective, the objective of the studies by Taibi and Lenarduzzi~\cite{microservice-bad-smells-m5}, by Bogner et al.~\cite{collaborative-repo-antipatterns}, and by Carrasco et al.~\cite{migrating-towards-microservices-m13} is much closer to ours.
Taibi and Lenarduzzi~\cite{microservice-bad-smells-m5} report on \abb{a} survey submitted to practictioners experienced with microservices.
The survey allowed Taibi and Lenarduzzi to identify 11 microservice-specific architectural smells, each with a refactoring solution allowing to \js{resolv}e it.
Of such smells and refactorings, only \js{four} can be related to the design principles of microservices pertaining to the process viewpoint (see Table~\ref{tab:antipatterns-coverage}).
By integrating the work by Taibi and Lenarduzzi with other carefully selected white/grey literature, we managed to extend the set of architectural smells and refactorings pertaining to the process viewpoint with \js{three} additional smells and \js{ten} additional refactorings. 

Bogner et al.~\cite{collaborative-repo-antipatterns} present a systematic literature review identifying and documenting architectural smells in SOA-based architectural styles, including microservices. 
Altough the main focus of their review is on the broader SOA, several smells apply also to microservices.
However, the review by Bogner et al.~\cite{collaborative-repo-antipatterns} differs from ours, as it focuses only on white literature, and since it does not discuss the architectural refactorings allowing to \js{resolv}e the identified smells.

Carrasco et al.~\cite{migrating-towards-microservices-m13} systematically analyses the white and grey literature on architectural smells that can occur while migrating from monoliths to microservice-based solutions.
They present \js{nine} common smells with their potential solutions, which all pertain to the actual development and operation of microservice-based applications (i.e.,~development and physical viewpoints).
The study by Carrasco et al.~\cite{migrating-towards-microservices-m13} hence differs from ours, as we focus on the dynamic aspects of microservices that interact at runtime (i.e.,~process viewpoint).

In summary, to the best of our knowledge, there is currently no study classifying the architectural smells possibly violating the design principles of microservices pertaining to the process viewpoint, together with the refactorings that permit \js{resolv}ing such smells.
This is the scope of our study, which we have presented in this paper.

\section{Conclusions}
\label{sec:conclusions}
%recap of what presented
We presented the results of a multivocal review focused on identifying architectural smells indicating possible violations of the independent deployability, horizontal scalability, fault isolation and decentralisation of microservices, as well as the refactorings allowing to \js{resolv}e such smells.
More precisely, we presented a taxonomy organising \js{seven} architectural smells and 16 refactorings, by associating each smell with the design principle(s) it violates, and each refactoring with the smell it \js{resolv}es.
We then provided an overview of the actual recognition of such smells and refactorings in the selected literature.
\ab{We} also discussed why each architectural smell violates the design principle it pertains to, and how each architectural refactoring allows \js{resolving} its corresponding smell.

%importance of what presented for researchers and practitioners
We believe that our study can be of help to both researchers and practitioners interested in microservices.
Together with the review by Carrasco et al.~\cite{migrating-towards-microservices-m13}, our results can help them to understand the well-known architectural smells for microservices, and to choose among the refactorings allowing to \js{resolv}e such smells. 
This can have a pragmatic value for practitioners, \js{who can exploit the results of our study in their daily work with microservices}. % as a starting point for experimenting microservices, as well as for their daily work with microservices.
%\marginpar{repeat?}
It can also help researchers to shape new solutions and to establish future research directions. 

% future work (development of a design-time support for automatically identifying architectural smells and for \js{resolv}ing them)
%In this direction, we
\abb{We} plan to exploit our results to develop a design-time support for eliminating architectural smells from microservice-based applications.
Our idea is to exploit existing languages for the specification of microservice-based applications (such as TOSCA~\cite{tosca}, for instance).
We then plan to develop a tool for processing the specification of a microservice-based application, to automatically detect the architectural smells occurring in such application, and to suggest the architectural refactorings \js{resolving} such smells.

\begin{acknowledgements}
    This work \abb{was} partly funded by the POR-FSE project \textit{AMaCA} (Regione Toscana), and by the project \textit{DECLware} (PRA\_2018\_66, University of Pisa).
    
    \smallskip \noindent
    \textbf{Published version} Neri D, Soldani J, Zimmermann O, Brogi A. \textit{Design principles, architectural smells and refactorings for microservices}, SICS Software-Intensive Cyber-Physical Systems (2019). DOI: 10.1007/s00450-019-00407-8   
\end{acknowledgements}

{
\footnotesize
All links were last followed on March 11th, 2019.
}

\end{document}